\def\year{2021}\relax
\documentclass[letterpaper]{article} % DO NOT CHANGE THIS
\usepackage{aaai21}  % DO NOT CHANGE THIS
\usepackage{times}  % DO NOT CHANGE THIS
\usepackage{helvet} % DO NOT CHANGE THIS
\usepackage{courier}  % DO NOT CHANGE THIS
\usepackage[hyphens]{url}  % DO NOT CHANGE THIS
\usepackage{graphicx} % DO NOT CHANGE THIS
\usepackage{natbib}  % DO NOT CHANGE THIS AND DO NOT ADD ANY OPTIONS TO IT
\usepackage{caption} % DO NOT CHANGE THIS AND DO NOT ADD ANY OPTIONS TO IT
\usepackage{physics}
\usepackage{amsmath}
\usepackage{tikz}
\usepackage{mathdots}
\usepackage{yhmath}
\usepackage{cancel}
\usepackage{color}
\usepackage{siunitx}
\usepackage{array}
\usepackage{multirow}
\usepackage{amssymb}
\usepackage{gensymb}
\usepackage{tabularx}
\usepackage{extarrows}
\usepackage{booktabs}
\usetikzlibrary{fadings}
\usetikzlibrary{patterns}
\usetikzlibrary{shadows.blur}
\usetikzlibrary{shapes}
\usepackage{hyperref}
\usepackage{listings}
\title{Human-guided Collaborative Problem Solving:\\ A Natural Language based Framework}
\author{
    Harsha Kokel,\textsuperscript{\rm 1} 
    Mayukh Das,\textsuperscript{\rm 1,2} 
    Rakibul Islam,\textsuperscript{\rm 3}  
    Julia Bonn,\textsuperscript{\rm 4} 
    Jon Cai,\textsuperscript{\rm 4} 
    Soham Dan,\textsuperscript{\rm 5} \\
    Anjali Narayan-Chen,\textsuperscript{\rm 6} 
    Prashant Jayannavar,\textsuperscript{\rm 6} 
    Janardhan Rao Doppa,\textsuperscript{\rm 3} 
    Julia Hockenmaier,\textsuperscript{\rm 6} \\
    Sriraam Natarajan,\textsuperscript{\rm 1} 
    Martha Palmer,\textsuperscript{\rm 4}
    Dan Roth \textsuperscript{\rm 5}

}
\affiliations {
\textsuperscript{\rm 1} University of Texas at Dallas,
\textsuperscript{\rm 2} Microsoft Research India,
\textsuperscript{\rm 3}  Washington State University, 
\textsuperscript{\rm 4}  University of Colorado Boulder, \\
\textsuperscript{\rm 5}  University of Pennsylvania, 
\textsuperscript{\rm 6}  University of Illinois at Urbana-Champaign \\
}
\begin{document}

\maketitle

\begin{abstract}
We consider the problem of human-machine collaborative problem solving as a planning task coupled with natural language communication. 
Our framework consists of three components -- a natural language engine that parses the language utterances to a formal representation and vice-versa, a concept learner that induces 
generalized concepts for plans based on limited interactions with the user and an HTN planner that solves the task based on human interaction. 
We 
illustrate the 
ability
of this framework 
to address the key challenges of collaborative problem solving by demonstrating it 
on a collaborative building task in a Minecraft-based blocksworld domain. 
The accompanied demo video is available at \href{https://starling.utdallas.edu/papers/collaborative-ps}{https://starling.utdallas.edu/papers/collaborative-ps}.
\end{abstract}
\vspace{-0.5em}
\section{Introduction}

It is well known that human-machine collaborative planning and problem solving is quite challenging as it requires shared perception of the world, sophisticated language understanding, fluent execution, bi-directional communication and contextual understanding. As part of the DARPA Communicating with Computers program, we consider the problem of human-machine collaboration in
a Minecraft environment with dynamics similar to BlocksWorld.
Our task is to build the target structures in a stipulated build region. 
Two players, an architect and a builder, collaborate and communicate using natural language interaction via chat interface. The architect, played by human, has access to the target structure and can see the current state in build region. The builder, the machine, can move in the build region and place/remove blocks. Since the builder does not have access to target structure, there is a need for the architect to describe the structure to the builder via chat interface. 

\begin{figure}[!t]
\centering
\includegraphics[width=\columnwidth]{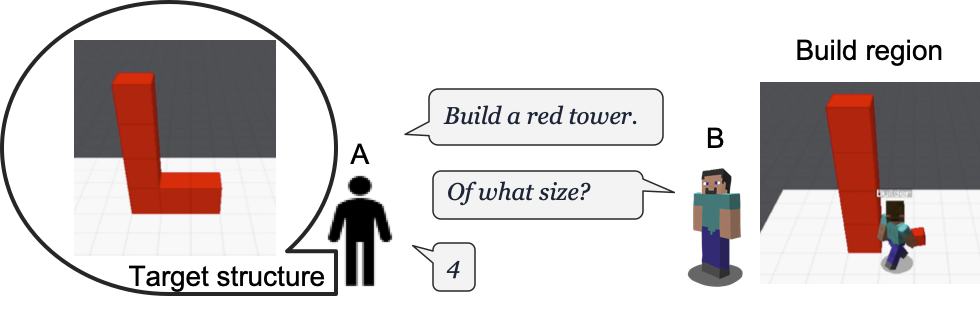}  
\caption{\small Target structure on the left is visible only to the architect (A). Architect instructs the builder (B) to build a red tower. B seeks clarification about the size and then proceeds to build the tower in the build region.}
\label{fig:illustration}
\vspace{-1em}
\end{figure}

For a successful target structure construction, the architect must
decompose the target structure to smaller structures and instruct the builder to achieve those subtasks. The builder must parse the instruction and achieve the subtask. The conundrums posed by the Minecraft-based blocks world task are \textbf{1.} the communication between the architect and the builder is inherently bi-directional (see for example Figure \ref{fig:illustration}), \textbf{2.} the builder should be able to seek clarifications as required, \textbf{3.}  both players must share some initial structures in the vocabulary and expand the vocabulary with experience. 
These problems of the proposed task highlight the key challenges of the collaborative planning problem: {\em bi-directional communication, contextual understanding, composable vocabulary and a powerful concept learner that can induce new, rich concepts based on limited interaction and experience}. We demonstrate our collaborative planning and problem-solving agent that addresses these key challenges. Our system has the capability to understand, quantify and measure ``what-it-doesn't-know" (dearth of relevant information) and leverage that understanding to elicit ``advice/knowledge" at the most appropriate decision points from the human and potentially learn better plans for increasingly complex structures. Some recent works (for e.g. Narayan-Chen et al. \citeyear{Narayan-ChenJH19} and K\"ohn et al. \citeyear{kohn-etal-2020-mc}) introduced a similar Minecraft environment, but focused on the dialogue generation and instruction giving; instead of dialogue understanding, concept induction, and planning challenges. 

\vspace{-1em}

\section{System}

For this collaborative building task, the initial knowledge of both the players include few colors, directions, primitive structures and block indicators (as inductive bias for the system). Blocks are limited to six colors: red, blue, green, purple, orange, and yellow. For simplicity, we only use the directions w.r.t the architect's view point.  Eight primitive structures include block, tower, row, column, square, rectangle, cube, and cuboid. 
Block indicators include guides for pointing to a single block that is a component of the structure. 
For example, the bottom-end of tower. Architect and the builder take turns on the chat window. 

The game begins with a simple target structure in the oracle window and a greeting from the builder to the architect. Architect responds with an instruction. The architect can instruct the builder to construct a structure or trigger UNDO command. 
If needed builder elicits required information from the architect by asking question (see for e.g. Fig. \ref{fig:illustration}). 
When a structure is build, builder would offer to remember the build structure for reuse. The architect can then name the structure and describe its height, width and length for the builder. The builder might ask some yes/no questions to induce a generalized concept of the structure~\cite{goci}.  If the builder is successful, the new structure is then added to builder's capabilities and can be treated as a primitive structure. The game proceeds to a new session, where the complexity of the target structure is increased.
\vspace{-0.5em}
\subsection{Builder Framework}

Our builder architecture has four main components.

\noindent
\textbf{1. Minecraft Simulator:} Minecraft is a game where avatars can navigate in a complex 3D world and place/remove blocks to build structures. Project Malmo \cite{JohnsonHHB16}, built on top of Minecraft, exposes an API for flexible AI experiments. We extended it for our collaborative building task by adding dialogue manager and agent control module. The \textbf{dialogue manager} triages the chat messages from the architect as build instruction, clarification, or concept explanation. It accordingly triggers a request to NLP engine, planner, concept learner, or agent-control module. The dialogue manager can also send message to the architect. The \textbf{agent control} module interacts with MALMO API to send action commands to the Minecraft environment.

\noindent
\textbf{2. NLP Engine:} We employed two natural language parsers to parse and translate the chat messages to predicate logic representation. The \textbf{rule-based parser} uses predefined templates for translation of simpler sentences while the \textbf{AMR parser} supports free form sentences of varying complexity. We use our Spatial-AMR \cite{bonn-etal-2020-spatial}, an extention of the Abstract Meaning Representation (AMR) annotation schema, which enables more expressive representation for spatial relationships required in the three-dimensional domain of Minecraft. Our AMR to logic translator is based on the theoretical framework from \citet{johanbos}.

\noindent
\textbf{3. Planner:}
This module is responsible for providing the action sequence of placing blocks in  Minecraft. The \textbf{repository} contains the description of various structures. Initially the repository only contains the 8 primitive structures, but it is gradually expanded with human interaction.
The predicate logic representation of build instructions is first examined for completeness by the \textbf{logic parser}. It identifies what information is missing, if any, and seeks it from the architect. If the instruction is complete, it converts the logical representation to the HTN format \cite{nau1999shop} and triggers the \textbf{JSHOP2} planner \cite{ilghami2006documentation} to seek the build plan. This plan is then returned to the dialogue manger. An independent investigation confirms that the HTN planner is better suited than classical planner for the Minecraft building task \cite{wichlacz2019construction}. 

\noindent
\textbf{4. Concept Learner:}
For effective human-machine collaborative problem solving, it is important that the machine learns new structural concepts as instructed by the human. The machine generalizes structural concepts from one or few instances described by the human architect and enhances its knowledge \textit{repository} progressively. For this, we employ our Guided One-shot Concept Induction (GOCI) framework \cite{goci}, which learns decomposable concepts that can be represented as conjunction of other concepts. For example, the structure shaped like an ``L" can be composed as combination of a tower and a row. The structure learned by GOCI are appended to the repository in the planner for further use. 
This gives the builder a notion of a cognitively intelligent agent that learns new concepts by collaborating with human architects and becomes capable of building increasingly complex structure.

\textbf{Discussion:} We presented our Minecraft based BlocksWorld domain where the planner and the human solve the tasks sequentially. Our system exploits human-machine communication to effectively solve tasks using a small number of human interactions. The key aspect of our system is that the tasks can be saved and reused based on our concept induction method. This allows for enriching the set of plans by building upon previously learned concepts while also enriching the vocabulary of communication with the users. The use of predicate logic notation (specifically horn clauses) allows for the system to explain the induced concepts as simple if then rules, thus increasing the explainability of the system's choices while increasing the trust of the user. Future directions include comprehensive evaluation allowing  for human errors in definitions and feedback and extending the framework to noisy, stochastic worlds where the concepts are represented by a richer probabilistic logic framework.
 
%\clearpage
\footnotesize
\textbf{Acknowledgments: } We gratefully acknowledge the support of CwC Program Contract W911NF-15-1-0461 with US Defense Advanced Research Projects Agency (DARPA) and Army Research Office (ARO). We do not reflect views of the DARPA, ARO or the US government 
\vspace{-1em}

\bibliography{biblio}

\clearpage

\end{document}